\documentclass[amsmath, amssymb, aps, prx, twocolumn]{revtex4-2}

\usepackage{graphicx}
\usepackage{dcolumn}
\usepackage{bm}
\usepackage{array}
\usepackage{float}
\usepackage{mathrsfs}
\usepackage{multirow}
\usepackage{varwidth}
\usepackage{amsmath}
\usepackage{amssymb}
\usepackage{stmaryrd}
\usepackage{graphicx}
\usepackage{hyperref}
\usepackage{subcaption}
\usepackage{cancel}
\usepackage{algorithm}
\usepackage{algorithmic}
\usepackage{booktabs}
\usepackage{makecell}
\usepackage{amsthm}

\usepackage{caption}

\newtheorem{theorem}{Theorem}
\newtheorem{observation}{Observation}

\begin{document}

\title{Hamiltonian-Informed Point Group Symmetry-Respecting Ansatz for Variational Quantum Eigensolver}
\author{Runhong He$^{1}$, Arapat Ablimit$^{2}$, Xin Hong$^{1}$, Qiaozhen Chai$^{1}$, Junyuan Zhou$^{3}$, Ji Guan$^{1}$, Guolong Cui$^{4}$}
\author{Shenggang Ying $^{1}$}
\email{yingsg@ios.ac.cn}
\address{
1. Key Laboratory of System Software (Chinese Academy of Sciences), Institute of Software, Chinese Academy of Sciences, Beijing 100190, China.\\
2. College of Physical Science and Technology, Xinjiang University, Urumgi 830017, China.\\
3. MindSpore Quantum Special Interest Group.\\
4. Arclight Quantum Co., LTD. Chinese Academy of Sciences, Beijing 101408, China.\\
}
\date{\today}

\begin{abstract}
Solving molecular energy levels via the Variational Quantum Eigensolver
(VQE) algorithm represents one of the most promising applications
for demonstrating practically meaningful quantum advantage in the
noisy intermediate-scale quantum (NISQ) era. To strike a balance between
ansatz complexity and computational stability in VQE calculations,
we propose the HiUCCSD, a novel symmetry-respecting ansatz engineered
from the intrinsic information of the Hamiltonian. We theoretically
prove the effectiveness of HiUCCSD within the scope of Abelian point
groups. Furthermore, we compare the performance of HiUCCSD and the
established SymUCCSD \href{https://journals.aps.org/pra/abstract/10.1103/PhysRevA.105.062452}{[Phys. Rev. A $\mathbf{105}$,}  \href{https://journals.aps.org/pra/abstract/10.1103/PhysRevA.105.062452}{062452 (2022)]} via VQE and
Adaptive Derivative-Assembled Pseudo-Trotter (ADAPT)-VQE numerical
experiments on ten molecules with distinct point groups. The results
show that HiUCCSD achieves equivalent performance to SymUCCSD for
Abelian point group molecules, while avoiding the potential performance
failure of SymUCCSD in the case of non-Abelian point group molecules.
Across the studied molecular systems, HiUCCSD cuts the parameter count by 18\%–83\% for VQE and reduces the excitation operator pool size by 27\%–84\% for ADAPT-VQE, as compared with the UCCSD ansatz. With enhanced
robustness and broader applicability, HiUCCSD offers a new ansatz
option for advancing large-scale molecular VQE implementation.

\end{abstract}

\maketitle

\section{Introduction}\label{sec:Introduction}
The properties of matter are fundamentally determined by molecular
energy levels \cite{molecular_property_calculations}. Knowledge of
these energy levels enables the location of stable structures, the
prediction of reaction rates, and the determination of optical properties,
etc. Within classical computational frameworks, molecular energies
can be exactly determined through the Full Configuration Interaction
(FCI) method \cite{review_molecular_calculations}. However, the FCI
method's computational cost scales exponentially with system size,
making it feasible only for small molecular systems \cite{szabo}.
For larger molecular systems, approximations have to be introduced
to balance computational cost and accuracy \cite{density_functional_theory,DFT}.
A prominent example is the Coupled-Cluster Singles and Doubles (CCSD)
method \cite{CCSD}, which truncates the cluster operator at single-
and double-excitations, significantly reducing computational complexity.
However, in strongly correlated systems, CCSD generally fails to yield
accurate correlation energies.

Quantum algorithms that exploit the inherent properties of quantum
superposition and entanglement hold the promise of achieving substantial
speedups over state-of-the-art classical algorithms for solving a
class of computationally challenging problems \cite{qc_nielsen}.
The Quantum Phase Estimation (QPE) \cite{QPE} algorithm was the first
quantum framework proven to efficiently tackle Hamiltonian eigenvalue
problems, a core task in quantum chemistry and condensed matter physics.
However, QPE’s stringent hardware requirements --- including robust
quantum error correction and extended coherence times --- currently
surpass the performance limits of near-term Noisy Intermediate-Scale
Quantum (NISQ) devices \cite{NISQ,NISQ_1}, hindering its practical
deployment in the current quantum technology landscape.

In recent years, the Variational Quantum Eigensolver (VQE) algorithm
\cite{VQE,VQE_first,vqe_review,vqe_review2} has garnered widespread
attention due to its adoption of a quantum-classical hybrid architecture,
which is inherently compatible with NISQ constraints. During VQE execution,
a quantum processor implements a parameterized quantum circuit (termed
an ansatz) to prepare trial quantum states, while a classical computer
computes the energy expectation value (serving as the loss function)
and associated gradients from quantum measurement results. These classical
processing outputs are then used to iteratively update the ansatz
parameters via classical optimization, enabling progressively more
accurate approximations of the target Hamiltonian’s ground-state energy.
Compared to QPE, VQE is distinguished by its shallow circuit depth
and reduced coherence time requirements, rendering it a promising
candidate for practical implementation in the NISQ era.

A critical determinant of VQE’s performance lies in the design of
the ansatz. In contrast to hardware-efficient ansatzes \cite{HEA_nature}, which are readily implementable but prone to parameter optimization
challenges (e.g., barren plateaus \cite{Barren_plateaus}) as system
size scales, the Unitary Coupled Cluster with Single and Double (UCCSD)
\cite{VQE_first,ucc_review} ansatz --- a unitary extension of the
classical CCSD --- has garnered extensive attention and in-depth
investigation due to its clear physical interpretation and robust
convergence. Notably, the UCCSD ansatz is typically constructed in
a problem-agnostic manner: it does not incorporate the specific structural
or electronic properties of the target system, thereby introducing
numerous redundant excitation operators \cite{VQE_point_group_symmetry}.
In general, for a system with $N$ molecular spin-orbitals, the number
of excitations in UCCSD scales as $O(N^{4})$. This high computational
complexity severely limits UCCSD’s practical applicability to medium-to-large
molecular systems. 

To address this scalability bottleneck, researchers have developed
a range of UCCSD variants, each navigating distinct trade-offs between
expressiveness, structural simplicity, and generality. For instance,
Ref. \cite{k-UpCCGSD} proposed the $k$-Unitary Pair Coupled-Cluster
with Generalized Singles and Doubles ($k$-UpCCGSD) ansatz, which
reduces circuit complexity by constraining the excitation scope while
preserving key electron correlation effects. However, its performance
is highly system-dependent: an improper choice of $k$ can lead to
under-parameterization (compromised expressiveness) or over-parameterization
(unnecessary quantum resource consumption).

The Adaptive Derivative-Assembled Pseudo-Trotter (ADAPT)-VQE algorithm
\cite{ADAPT-VQE,ADAPT_VQE_QEB,ADAPT_VQE_ES,ADAPT_VQE_Pruned,ADAPT_VQE_CEO}
offers a dynamic solution to the ansatz design challenge. Starting
from a minimal initial ansatz (e.g., the Hartree-Fock state), it iteratively
incorporates excitation operators from a predefined ``operator pool''
that maximize the energy gradient magnitude, thereby adaptively constructing
a compact ansatz tailored to the specific target system. Furthermore,
ADAPT-VQE has been demonstrated to mitigate optimization challenges
associated with barren plateaus and local minima \cite{ADAPT_VQE_vs_bp}.
While ADAPT-VQE effectively eliminates redundant operators, it introduces
non-trivial computational overhead: each iteration requires computing
the gradient vector for an operator pool of size $O(N^{4})$, which
substantially increases the algorithm’s time complexity and quantum-classical
communication costs. Thus, reducing the size of the operator pool
has emerged as a crucial research direction for enhancing the practical
applicability of ADAPT-VQE \cite{ADAPT_VQE_qubit,ADAPT_VQE_pauli_linear_pool}.

The Symmetry-constrained UCCSD (SymUCCSD) ansatz \cite{VQE_point_group_symmetry}
addresses the redundancy in the UCCSD operator pool by leveraging
the fundamental principle of molecular symmetry --- a ubiquitous
property of chemical systems that has long guided classical quantum
chemistry. For excitations contributing to the ground-state energy,
their corresponding excited states must preserve the symmetry of the
reference state (i.e., maintain the same irreducible representation,
irrep). This theoretical insight enables SymUCCSD to systematically
prune redundant operators that fail to satisfy symmetry constraints,
drastically reducing the operator pool size without compromising expressiveness.

Despite its successes, SymUCCSD faces several limitations that impede
its widespread adoption in broader quantum chemistry applications.
First, point group identification is non-trivial: determining a molecule’s
correct point group requires precise knowledge of its geometry and
thorough understanding of group theory. For flexible molecules (e.g.,
proteins, conformational isomers) or systems with dynamic geometries
(e.g., reaction intermediates), point group assignment may become
ambiguous or computationally expensive, as it necessitates first optimizing
the molecular structure to a symmetric conformation. Second, SymUCCSD’s
theoretical framework is restricted to Abelian point groups, which
poses a critical constraint given that many chemically important molecules
belong to non-Abelian point groups (e.g., methane, CH$_{4}$ ($\mathrm{T_d}$);
ammonia, NH$_{3}$ ($C_{3\text{v}}$)). A common workaround for non-Abelian
systems is to downgrade them to an Abelian subgroup for compatibility
with SymUCCSD, but this simplification may lead to incorrect screening
of excitations --- either removing critical energy contributing operators
(resulting in loss of expressiveness) or retaining redundant operators
(wasting quantum resources). Such issues not only undermine the efficiency
of the operator pool pruning but also propagate to the ground-state
energy calculation, leading to insufficient accuracy.

In this paper, we propose a systematic and robust excitation screening
method. Specifically, we prove that for Abelian point groups, the
electronic integrals in the Hamiltonian will vanish if the corresponding
excitation operators do not satisfy the molecular point group symmetry
requirements. Based on this theoretical result, we can eliminate symmetry-violating
excitation operators from a predefined ansatz by checking the values
of their corresponding electronic integrals in the Hamiltonian, thereby
constructing a compact, symmetry-respecting ansatz. We refer to this
method as Hamiltonian-Informed UCCSD (HiUCCSD). HiUCCSD eliminates
the need to explicitly consider molecular configurations or identify
point group symmetry, thereby avoiding result distortion caused by
incorrect point group assignment. Since the electronic integrals in
the Hamiltonian have been precomputed, the execution of this algorithm
is nearly cost-free. For molecules belonging to Abelian point groups,
HiUCCSD delivers the same performance outcomes as SymUCCSD. Notably,
numerical results demonstrate that HiUCCSD can also be extended to
molecules with non-Abelian point groups, enabling its application
to a significantly broader range of molecular systems.

The structure of the remainder of this paper is organized as follows:
In Section \ref{sec:Methods}, we introduce the fundamentals of the
VQE algorithm (\ref{subsec:VQE-Algorithm}), point group symmetry
and SymUCCSD (\ref{subsec:Point-Group-Symmetries}), along with the
theoretical underpinnings of the HiUCCSD (\ref{subsec:HiUCCSD}).
In Section \ref{sec:Numerical-Results}, we perform numerical experiments
with VQE (\ref{subsec:VQE-Implementation}) and ADAPT-VQE (\ref{subsec:ADAPT-VQE-Implementation}),
aiming to compare and analyze the performance of SymUCCSD and HiUCCSD.
In Section \ref{sec:Conclusion}, we present the conclusions.

\section{Methods }\label{sec:Methods}

\subsection{VQE Algorithm}\label{subsec:VQE-Algorithm}

We use indices $i,j,k,\dots$ to label occupied spin orbitals, $a,b,c,\dots$
to label virtual (unoccupied) spin orbitals, and $p,q,r,s$ to label
spin orbitals of either type. Additionally, we denote the total number
of spin orbitals as $N$, and the number of electrons as $n$. 

Within the Born-Oppenheimer approximation, where the nuclei of the
molecule are treated as motionless, the second-quantized electronic
Hamiltonian of a molecule in atomic units takes the form \cite{Quantum_computational_chemistry,szabo}
\begin{equation}
\mathscr{H}=\sum_{p,q}^{N}h_{q}^{p}\hat{a}_{p}^{\dagger}\hat{a}_{q}+\frac{1}{2}\sum_{p,q,r,s}^{N}h_{rs}^{pq}\hat{a}_{p}^{\dagger}\hat{a}_{q}^{\dagger}\hat{a}_{r}\hat{a}_{s},\label{eq:ham}
\end{equation}
where fermionic operators $\hat{a}_{i}^{\dag}$ (creation) and $\hat{a}_{i}$
(annihilation) obey anti-commutation relations:
\begin{equation}
\{\hat{a}_{i},\hat{a}_{j}^{\dag}\}=\delta_{i,j},\quad\{\hat{a}_{i},\hat{a}_{j}\}=\{\hat{a}_{i}^{\dag},\hat{a}_{j}^{\dag}\}=0.
\end{equation}
The one- and two-electron integrals $h_{q}^{p}$ and $h_{rs}^{pq}$
can be computed within specified basis sets \cite{szabo}, such as
the STO-3G basis set. Hamiltonian (\ref{eq:ham}) can be mapped to
Pauli strings using an encoding method, e.g., the Jordan-Wigner \cite{Jordan-Wigner},
and then to get the expectation value $E(\vec{\theta})=\langle\Psi(\vec{\theta})|\mathscr{H}|\Psi(\vec{\theta})\rangle$
in the trial state $|\Psi(\vec{\theta})\rangle$. The trial state
$|\Psi(\vec{\theta})\rangle$ is prepared from a reference state $|\Psi_{0}\rangle$
by a parameterized quantum circuit $U(\vec{\theta})$ with moderate
depth: $|\Psi(\vec{\theta})\rangle=U(\vec{\theta})|\Psi_{0}\rangle$.
The reference state is usually chosen to be the Hartree-Fock state. 

In the framework of VQE, an upper bound of the unknown ground-state
energy $E_{0}$ can be obtained by minizing the expectation value
of the Hamiltonian with respect to the variational parameters $\vec{\theta}$
through the Rayleigh-Ritz variational principle \cite{RRvariational}
\begin{equation}
\langle\Psi(\vec{\theta})|\mathscr{H}|\Psi(\vec{\theta})\rangle\geq E_{0}.
\end{equation}

The Unitary Coupled Cluster (UCC) is a chemistry-inspired ansatz commonly
employed to tackle quantum chemical problems \cite{ucc_review}. Within
UCC, the unitary operator of quantum circuit $U(\vec{\theta})$ takes
the form $U(\vec{\theta})=e^{T(\vec{\theta})-T^{\dagger}(\vec{\theta})}$,
where $T$ denotes any Hermitian excitation operator. To reduce the
depth of the quantum circuit, typically only single- and double-excitation
operators are retained, i.e., 
\begin{equation}
T=T_{1}+T_{2},
\end{equation}
where 
\begin{equation}
T_{1}(\vec{\theta})=\sum_{i,a}\hat{t}_{i}^{a}=\sum_{i,a}\theta_{i}^{a}\hat{a}_{a}^{\dagger}\hat{a}_{i},
\end{equation}
\begin{equation}
T_{2}(\vec{\theta})=\sum_{i>j,a>b}\hat{t}_{ij}^{ab}=\sum_{i>j,a>b}\theta_{ij}^{ab}\hat{a}_{a}^{\dagger}\hat{a}_{b}^{\dagger}\hat{a}_{i}\hat{a}_{j}.
\end{equation}
This variant of UCC is called UCCSD ansatz. Under first-order Trotter-Suzuki
approximation \cite{trotter}, the evolution operator can be decomposed
as:
\begin{equation}
e^{T(\vec{\theta})-T^{\dagger}(\vec{\theta})}\thickapprox\prod_{i,a}\text{e}^{\hat{t}_{i}^{a}-\hat{t}_{i}^{a\dagger}}\prod_{i>j,a>b}\text{e}^{\hat{t}_{ij}^{ab}-\hat{t}_{ij}^{ab\dagger}},
\end{equation}
This approximate ansatz is generally accepted in VQE because the variational
optimization can absorb most of the truncation errors \cite{VQE_trotter_one}.

The quantum circuit implementation of UCCSD ansatz has been extensively
investigated in the literature, featuring various overheads that include:
encodings \cite{Jordan-Wigner,Parity_basis,Bravyi-Kitaev} and efficient
circuit designs \cite{efficient_excitations_circ,ADAPT_VQE_CEO,ADAPT_VQE_QEB,adapt_9_cnot,exciation_9_cnot}.
In this paper, we adopt the method \cite{efficient_excitations_circ}
due to its low implementation cost: requiring only 2 and 13 CNOT gates
for single- and double-excitation operators, plus several single-qubit
rotation gates, respectively. 

\subsection{Point Group Symmetry and SymUCCSD}\label{subsec:Point-Group-Symmetries}

In this subsection, we briefly introduce the fundamentals of point
group symmetry, which underlies the principle of our method. For further
details on point group symmetry, readers are referred to Refs.\cite{group_theory_0,group_theory_1}.
To ground this discussion in a concrete context, we take the water
molecule (H$_{2}$O) as a paradigmatic example.

A point group refers to a set of symmetry operations that leave the
structure of an object (here, a molecule) unchanged before and after
their application. These symmetry operations include: Rotation ($C_{n}$,
rotation by $360^\circ/n$), Plane reflection ($\sigma$, reflection
across a specified plane), Inversion ($i$, inversion of all coordinates
about the center), Identity operation ($E$, the operation that leaves
the molecule unchanged), as well as their combinations (e.g. $S_{n}$,
generated by a $C_{n}$ rotation followed by a $\sigma$ reflection
in a plane perpendicular to the rotation axis). The term ``point''
underscores that all symmetry operations leave at least one point
in space invariant.

\begin{figure}[htbp]
\centering
\includegraphics[scale=0.012]{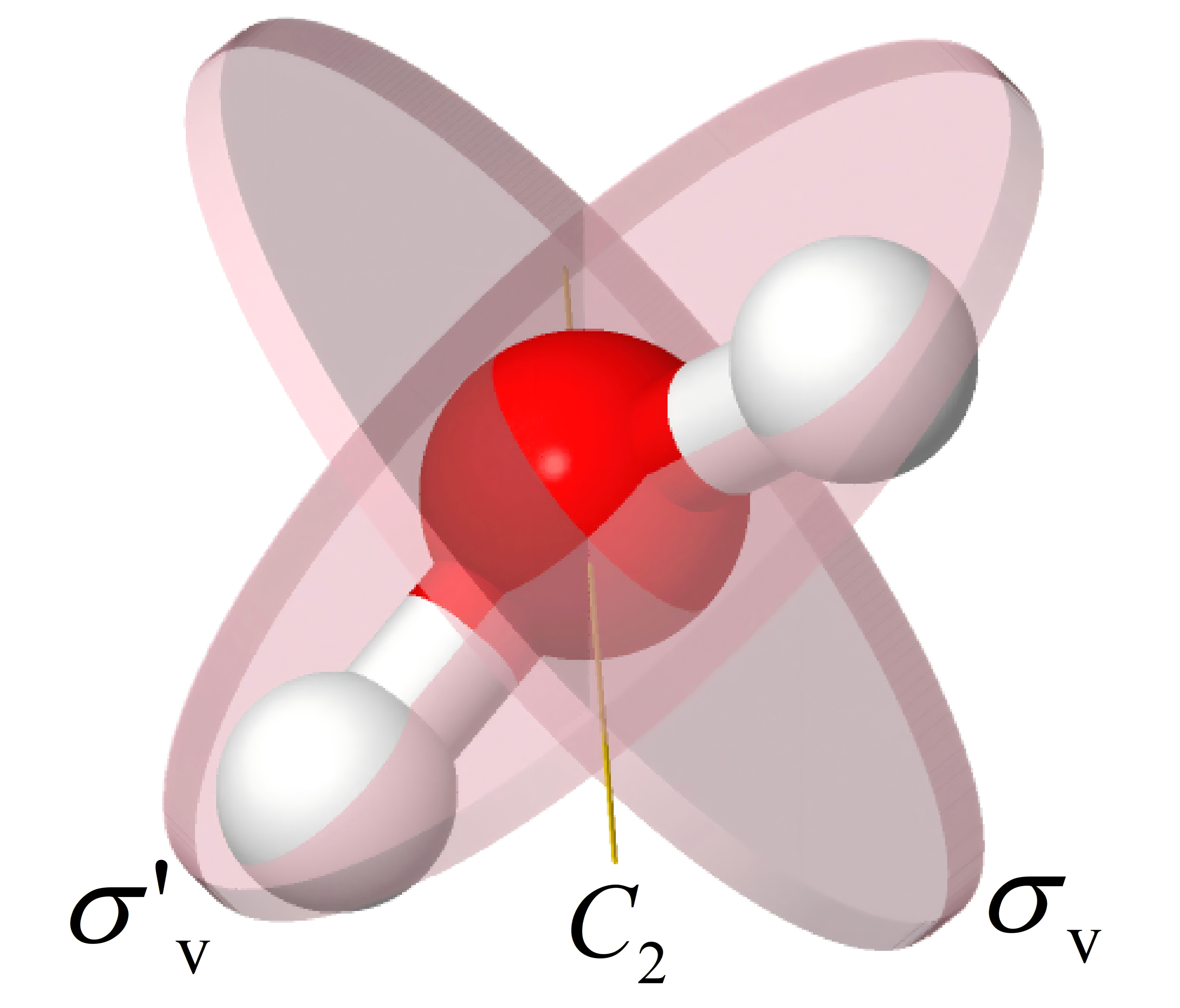}
\caption{The symmetry operations of the H$_2$O molecule (which belongs to the $C_{2\text{v}}$ point group) include the $C_{2}$ rotation and the $\sigma_{\text{v}}/\sigma'_{\text{v}}$ reflections.}
\label{fig_H2O}
\end{figure}

A molecule belongs to a point group if its spatial structure remains indistinguishable from the original after the application of any symmetry operation in that point group. For example, the water molecule ($\mathrm{H_2O}$) belongs to the $C_{2\text{v}}$ point group (see Fig.~\ref{fig_H2O}), as its symmetry comprises a $C_{2}$ rotational operation and two plane reflections through vertical planes, labeled $\sigma_{\text{v}}$ and $\sigma'_{\text{v}}$.

\begin{figure}[htbp]
 \centering
 
 \begin{subfigure}[b]{0.45\textwidth}
 \centering
\begin{tabular}{|c|c|c|c|c|}
\hline 
\textbf{$\boldsymbol{C}_{2\text{v}}$} & \textbf{$\boldsymbol{E}$} & $\boldsymbol{C}_{2}$ & \textbf{$\textbf{\ensuremath{\boldsymbol{\sigma}}}_{\textbf{v}}$} & \textbf{$\textbf{\ensuremath{\boldsymbol{\sigma}}}'_{\textbf{v}}$}\tabularnewline
\hline 
\hline 
\textbf{$\text{A}_{\text{1}}$} & 1 & 1 & 1 & 1\tabularnewline
\hline 
\textbf{$\text{A}_{\boldsymbol{2}}$} & 1 & 1 & -1 & -1\tabularnewline
\hline 
\textbf{$\boldsymbol{\text{B}}_{\boldsymbol{1}}$} & 1 & -1 & 1 & -1\tabularnewline
\hline 
\textbf{$\text{B}_{\boldsymbol{2}}$} & 1 & -1 & -1 & 1\tabularnewline
\hline 
\end{tabular}%
\caption{}
\end{subfigure}
\begin{subfigure}[b]{0.45\textwidth}
\centering
\begin{tabular}{|c|c|c|c|c|}
\hline 
\textbf{$\boldsymbol{C}_{2\text{v}}$} & \textbf{$\text{A}_{\boldsymbol{1}}$} & \textbf{$\text{A}_{\boldsymbol{2}}$} & \textbf{$\text{B}_{\boldsymbol{1}}$} & \textbf{$\text{B}_{\boldsymbol{2}}$}\tabularnewline
\hline 
\hline 
\textbf{$\boldsymbol{\text{A}}_{\boldsymbol{1}}$} & $\text{A}_{1}$ & $\text{A}_{2}$ & $\text{B}_{1}$ & $\text{B}_{2}$\tabularnewline
\hline 
\textbf{$\text{A}_{\boldsymbol{2}}$} & $\text{A}_{2}$ & $\text{A}_{1}$ & $\text{B}_{2}$ & $\text{B}_{1}$\tabularnewline
\hline 
\textbf{$\text{B}_{\boldsymbol{1}}$} & $\text{B}_{1}$ & $\text{B}_{2}$ & $\text{A}_{1}$ & $\text{A}_{2}$\tabularnewline
\hline 
\textbf{$\text{B}_{\boldsymbol{2}}$} & $\text{B}_{2}$ & $\text{B}_{1}$ & $\text{A}_{2}$ & $\text{A}_{1}$\tabularnewline
\hline 
\end{tabular}
\caption{}
\end{subfigure}

\caption{The character table (a) and the irrep direct product table (b) of
the $C_{2\text{v}}$ point group for the H$_{2}$O molecule. }
\label{fig:character_product_tables}
\end{figure}

Character tables and irreducible representation (irrep) direct product
tables are fundamental tools in the application of group theory  \cite{group_theory_0,group_theory_1}. A
character table summarizes the characters (traces of representation
matrices) of all irreps of a point group under each class of symmetry
operations. For simplicity, as adopted in Ref.~\cite{VQE_point_group_symmetry},
we restrict our discussion to Abelian point groups, for which the
characters of all symmetry operations are either $+1$ or $-1$. Fig.~\ref{fig:character_product_tables} (a) corresponds to the character
table of the $C_{2\text{v}}$ point group for the H$_{2}$O molecule.
The columns of Fig.~\ref{fig:character_product_tables} (a) denote
the conjugacy classes of symmetry operations inherent to $C_{2\text{v}}$:
$E$, $C_{2}$, and $\sigma_{\text{v}}/\sigma'_{\text{v}}$. The
rows represent the irreps of $C_{2\text{v}}$ ($\text{A}_{1}$, $\text{A}_{2}$,
$\text{B}_{1}$, $\text{B}_{2}$), all of which are one-dimensional
(hence their characters are scalar values). The numerical entries
in the table are the characters (i.e., the trace of the matrix representation)
of each irrep under the corresponding symmetry operation class. 
The letters $\text{A}$ and $\text{B}$ in the irrep labels serve to distinguish the symmetry of a molecular feature with respect to the $C_{2}$ rotation. 
Features labeled $\text{A}$ are symmetric (remaining unchanged) when
the molecule undergoes a $C_{2}$ rotation, while those marked $\text{B}$
are antisymmetric and altered by such a rotation.
Subscripts 1 and 2 further differentiate these features based on their behavior under two reflections. 
Specifically, subscript 1 denotes symmetry under the $\sigma_{\text{v}}$ reflection  while subscript 
2 denotes symmetry under the $\sigma'_{\text{v}}$ reflection.

The irrep product table, on the other hand, specifies the outcomes
of the direct product of any two irreps. Fig.~\ref{fig:character_product_tables}
(b) presents the irrep product table of the $C_{2\text{v}}$ point
group: rows and columns correspond to the four irreps of $C_{2\text{v}}$,
and each entry denotes the result of the direct product of the row
irrep and the column irrep. As an example, consider the direct product
of $\text{B}_{1}$ and $\text{B}_{2}$. The characters of $\text{B}_{1}$
are (1, -1, 1, -1), and those of $\text{B}_{2}$ are (1, -1, -1, 1).
Their direct product yields (1$\cdot$1, (-1)$\cdot$(-1), 1$\cdot$(-1),
(-1)$\cdot$1) = (1, 1, -1, -1), which correspond to the characters
of the $\text{A}_{2}$ irrep. 
\begin{figure}
\includegraphics[scale=0.026]{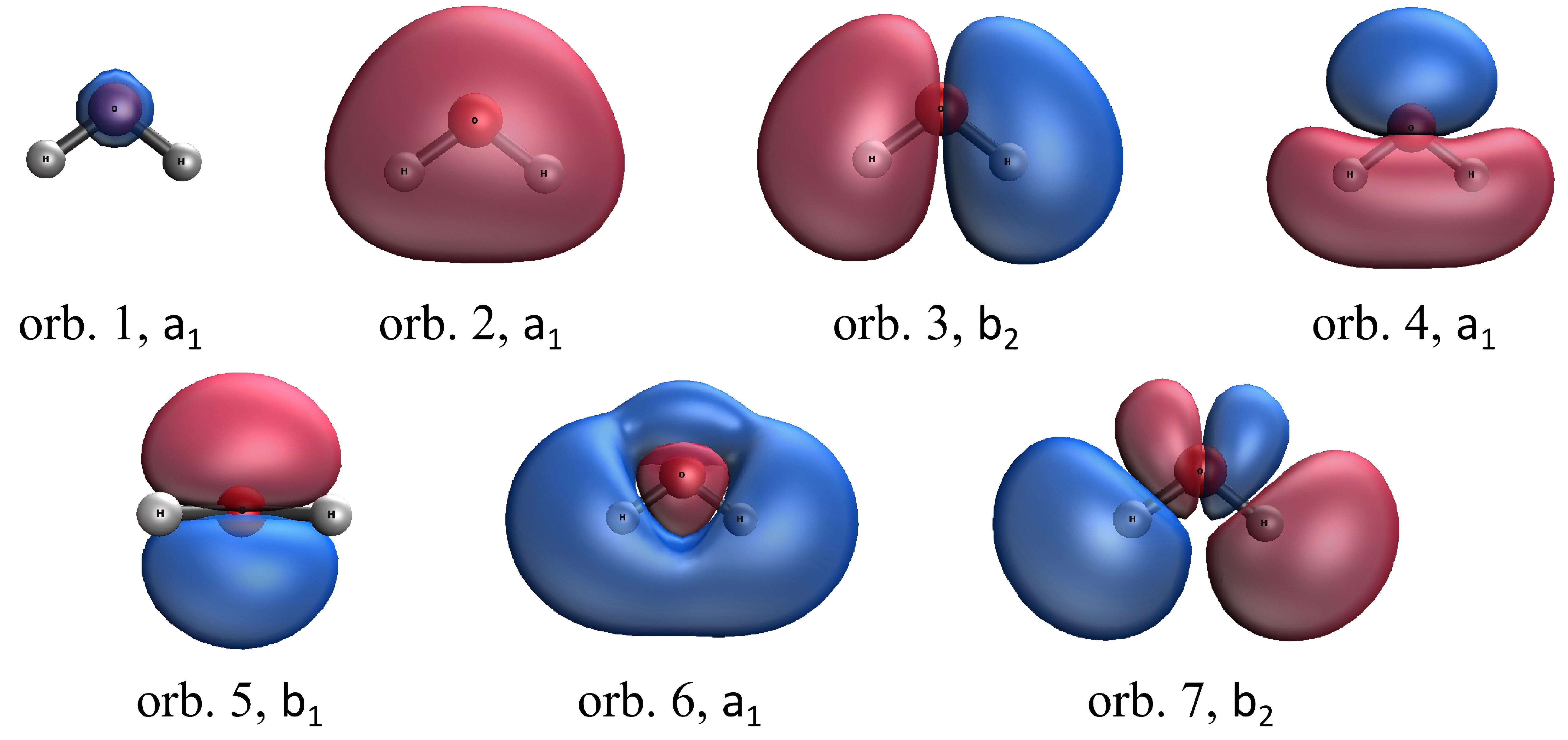}
\caption{Isosurfaces of the molecular orbital wavefunctions (isovalues $=$
0.08 arbitrary units). }
\label{fig:orbs}
\end{figure}

To illustrate spatial symmetry by combining it with the character
table, we present the molecular orbitals of H$_{2}$O. The isosurfaces
of these molecular orbitals are visualized in Fig.~\ref{fig:orbs},
where blue denotes positive values and red denotes negative values.
As shown in Fig.~\ref{fig:orbs}, the character of each irrep quantifies
the sign variation caused by specific symmetry operations. For instance,
for the $\text{b}_{2}$ orbitals (orb. 3 and orb. 7), the application
of either the $C_{2}$ rotation or $\sigma_{\text{v}}$ reflection
reverses the wavefunction’s sign (corresponding to a character of
-1), whereas the $\sigma'_{\text{v}}$ reflection (or $E$ operation)
leaves the sign unaltered (corresponding to a character of 1). In
contrast, for orbitals of the $\text{a}_{1}$ irrep (e.g., orb. 1),
the wavefunction remains unchanged regardless of the symmetry operation
applied, resulting in a consistent character of 1 across all operations.
Based on these character assignments, the direct product table of
irreps, as shown in Fig.~\ref{fig:character_product_tables} (b),
can be systematically constructed. Herein, we follow the conventional notation: the
irreps of molecular orbitals are denoted by lowercase letters (e.g.,
$\text{a}_{1}$), while those of quantum states are represented by
uppercase letters (e.g., $\text{A}_{1}$). 
\begin{figure}[htbp]
    \centering
    \begin{subfigure}[b]{0.15\textwidth}
        \centering
        \includegraphics[scale=0.04]{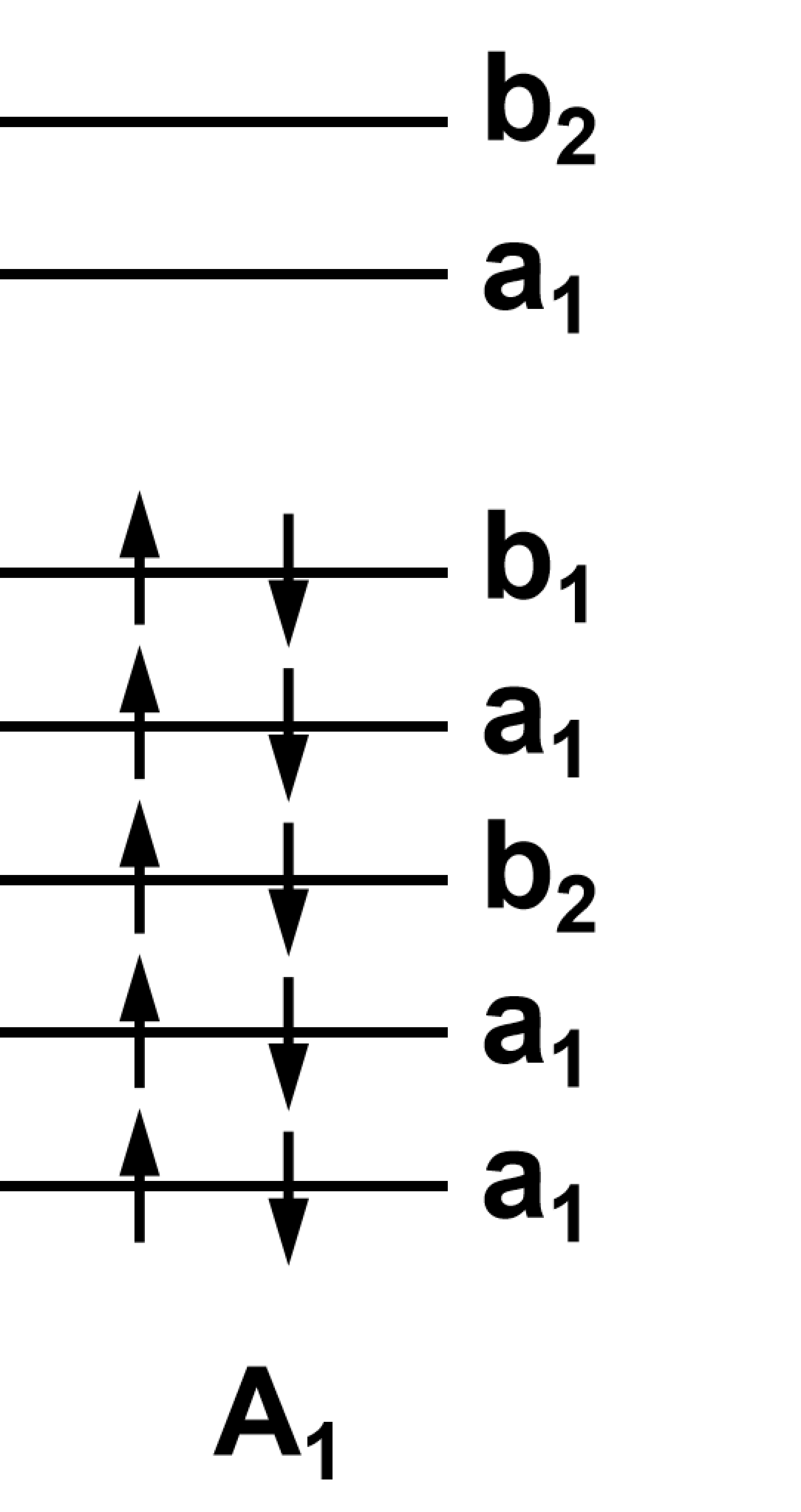}
        \caption{}
    \end{subfigure}
    \begin{subfigure}[b]{0.15\textwidth}
        \centering
        \includegraphics[scale=0.04]{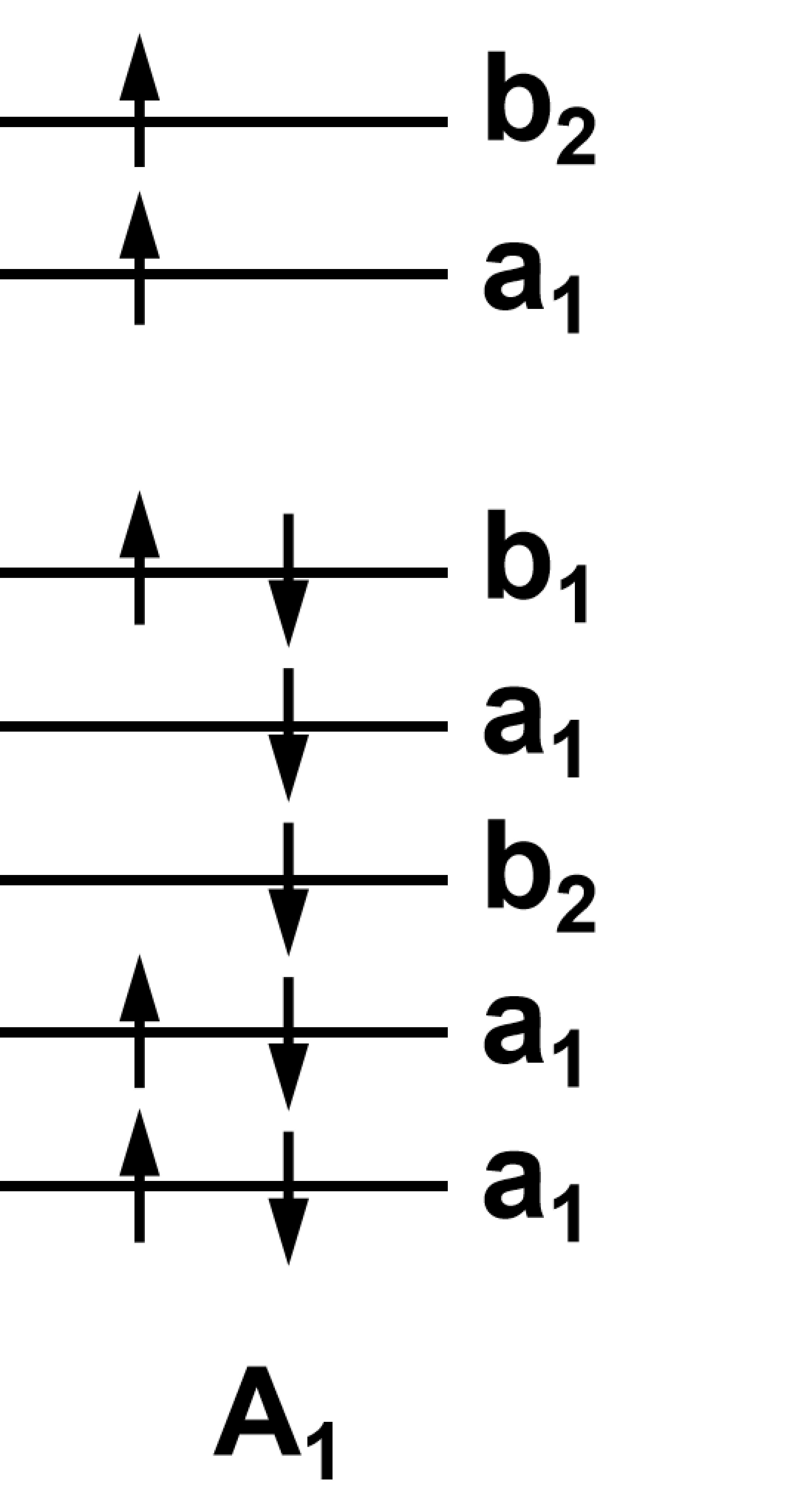}
        \caption{}
    \end{subfigure}
    \begin{subfigure}[b]{0.15\textwidth}
        \centering
        \includegraphics[scale=0.04]{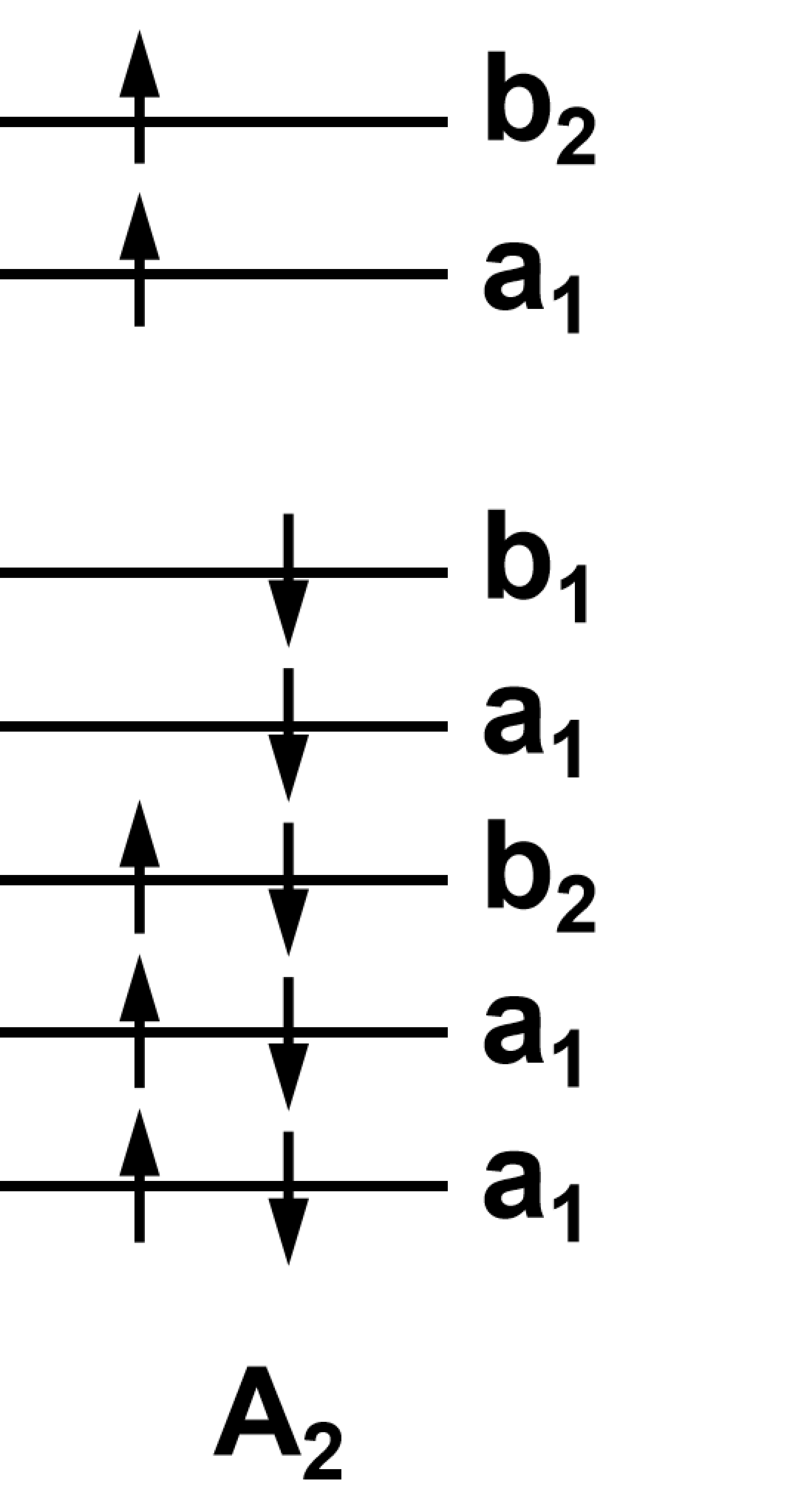}
        \caption{}
    \end{subfigure}
    \caption{Electron configurations of the $\mathrm{H_2O}$ molecule: (a) Reference state $|\Psi_{0}\rangle$ (Hartree-Fock state with $\mathrm{A}_1$ irrep); (b) Excited state $\hat{t}_{4,6}^{10,12}|\Psi_{0}\rangle$, which shares the same irrep with the reference state; (c) Excited state $\hat{t}_{6,8}^{10,12}|\Psi_{0}\rangle$ with an irrep distinct from that of the reference state. Here, the irrep of a molecular orbital is labeled in lowercase (e.g., $\mathrm{a_1}$), while that of a quantum state is denoted in uppercase (e.g., $\mathrm{A_1}$).}
    \label{fig:exaples}
\end{figure}

Next, we discuss the calculation of irreps for various molecular states
(including the Hartree-Fock state and single-excited Slater determinants),
to verify whether excitation operators meet the point group symmetry
requirements. The irrep of a molecular state is the same as the direct
product of the relevant molecular spin-orbitals.

We continue to take the H$_{2}$O molecule as the example. As shown in Fig.~\ref{fig:exaples},
three distinct electronic configurations are presented: the reference
state $|\Psi_{0}\rangle$ (the Hartree-Fock state), $\hat{t}_{4,6}^{10,12}|\Psi_{0}\rangle$
and $\hat{t}_{6,8}^{10,12}|\Psi_{0}\rangle$. The Hartree-Fock state
corresponds to a Slater determinant where all $n$ electrons occupy
the lowest-energy molecular orbitals, with their respective irreps
being $\text{a}_{1}$, $\text{a}_{1}$, $\text{b}_{2}$, $\text{a}_{1}$ and
$\text{b}_{1}$.

The irrep of the reference state $|\Psi_{0}\rangle$ is calculated
as 
\begin{equation}
\begin{split}
&(\mathrm{a}_{1}\!\otimes\!\mathrm{a}_{1})\!\otimes\!(\mathrm{a}_{1}\!\otimes\!\mathrm{a}_{1})\!\otimes\!(\mathrm{b}_{2}\!\otimes\!\mathrm{b}_{2})\!\otimes\!(\mathrm{a}_{1}\!\otimes\!\mathrm{a}_{1})\!\otimes\!(\mathrm{b}_{1}\!\otimes\!\mathrm{b}_{1}) \\
&=\mathrm{a}_{1}\otimes\mathrm{a}_{1}\otimes\mathrm{a}_{1}\otimes\mathrm{a}_{1}\otimes\mathrm{a}_{1}=\mathrm{A}_{1}.
\end{split}
\end{equation}
For the state $\hat{t}_{4,6}^{10,12}|\Psi_{0}\rangle$ illustrated
in Fig.~\ref{fig:exaples} (b), its irrep is
\begin{equation}
(\text{a}_{1}\!\otimes\!\text{a}_{1})\!\otimes\!(\text{a}_{1}\!\otimes\!\text{a}_{1})\!\otimes\!\text{b}_{2}\!\otimes\!\text{a}_{1}\!\otimes\!(\text{b}_{1}\!\otimes\!\text{b}_{1})\!\otimes\!\text{a}_{1}\!\otimes\!\text{b}_{2}=\text{A}_{1},
\end{equation}
 which matches that of the reference state. In contrast, the irrep
of the state $\hat{t}_{6,8}^{10,12}|\Psi_{0}\rangle$ (showed in Fig.~\ref{fig:exaples}(c)) is computed as
\begin{equation}
\begin{split}
&(\mathrm{a}_{1}\!\otimes\!\mathrm{a}_{1})\!\otimes\!(\mathrm{a}_{1}\!\otimes\!\mathrm{a}_{1})\!\otimes\!(\mathrm{b}_{2}\!\otimes\!\mathrm{b}_{2})\!\otimes\!\mathrm{a}_{1}\!\otimes\!\mathrm{b}_{1}\!\otimes\!\mathrm{a}_{1}\!\otimes\!\mathrm{b}_{2}\\
&=\mathrm{A}_{2}\neq\mathrm{A}_{1},
\end{split}
\end{equation}
indicating a mismatch with $|\Psi_{0}\rangle$. According to the conclusion
of SymUCCSD method \cite{VQE_point_group_symmetry}, only excited
states with the same irrep as the reference state contribute to the
energy. Therefore, the excited state $\hat{t}_{6,8}^{10,12}|\Psi_{0}\rangle$
does not satisfy the symmetry requirements, and the corresponding
excitation operator $\hat{t}_{6,8}^{10,12}$ should be removed from
the ansatz.

In a similar manner, we can traverse all excitation operators within
the UCCSD ansatz. For the H$_{2}$O molecule, the UCCSD ansatz initially
contains 140 excitation operators, but only 48 of them are retained,
as they preserve an irrep identical to that of the reference state.
This significant reduction in the number of retained operators greatly
lowers the implementation cost of the UCCSD ansatz.

\subsection{HiUCCSD }\label{subsec:HiUCCSD}
\begin{figure*}[htbp]
    \centering
    \renewcommand{\figurename}{Algorithm}
    \caption{Pseudocode of the HiUCCSD method for constructing symmetry-respecting ansatz.}
    \label{algo}
    \includegraphics[width=\textwidth]{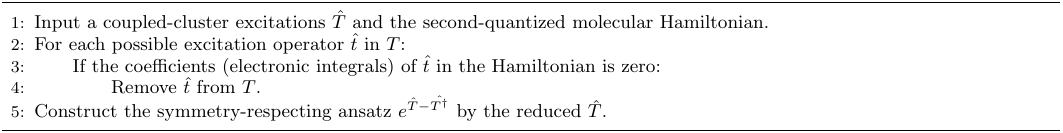}
    \renewcommand{\figurename}{Figure}
\end{figure*}







Here, we establish the theoretical foundation of our HiUCCSD method.
First, let us re-inspect Fig.~\ref{fig:exaples}, the following observation
can be readily deduced:

\begin{observation}
The irrep difference between the excited state and the reference state depends solely on the orbitals acted upon by the excitation operator. 
Specifically, if the direct product of the irreps of the orbitals for the creation part of the excitation operator matches that of the orbitals for the annihilation part, the excited state retains the reference state’s irrep; otherwise, their irreps differ.
\end{observation}

For example, in Fig.~\ref{fig:exaples} (b), the direct products of
the irreps of the orbitals acted upon by the creation and annihilation
parts of the excitation operator (i.e., $\hat{t}_{4,6}^{10,12}$)
are identical: $\text{b}_{2}\otimes\text{a}_{1}=\text{b}_{2}=\text{a}_{1}\otimes\text{b}_{2}$.
In contrast, for the excitation operator $\hat{t}_{6,8}^{10,12}$illustrated
in Fig.~\ref{fig:exaples} (c), the direct products of the irreps
of the orbitals targeted by its creation and annihilation parts are
distinct: $\text{b}_{2}\otimes\text{a}_{1}=\text{b}_{2}\neq\text{b}_{1}=\text{b}_{1}\otimes\text{a}_{1}$.
This discrepancy results in a change in the total irrep of the corresponding
excited state relative to the reference state.

Next, we formalize the following theorem, which constitutes the theoretical
cornerstone of the HiUCCSD method.

\begin{theorem}\label{theorm_1}
For any Abelian point group, the electronic
integrals in the Hamiltonian will vanish if the associated excitation
operators fail to satisfy the molecular point group symmetry requirements.
\end{theorem}

\begin{proof}
For any operation $R$ of the Abelian point group, the symmetry transformation
operator $\hat{\boldsymbol{O}}_{R}$ acts on a molecular spin orbitals
$\phi_{i}$ as follows:
\begin{equation}
\hat{\boldsymbol{O}}_{R}\phi_{i}=\gamma_{i}^{R}\phi_{i},
\end{equation}
where $\gamma_{i}^{R}$ denotes the irrep character of orbital $\phi_{i}$
under the symmetry operation $R$. For an Abelian group, this character
is always +1 or \textminus 1.

The Hamiltonian~(\ref{eq:ham}) of a molecular system includes one-electron
and two-electron integrals, which describe the core-electron interactions
and electron-electron Coulomb repulsion, respectively. These integrals
are formally defined as \cite{quantum_chemetry_book_2,szabo}:
\begin{equation}
h_{q}^{p}=\int d\mathbf{x}\phi_{p}^{*}(\mathbf{x})\left(-\frac{\nabla^{2}}{2}-\sum_{I}\frac{Z_{I}}{\left|\mathbf{r}-\mathbf{R}_{I}\right|}\right)\phi_{q}(\mathbf{x}),
\end{equation}
\begin{equation}
h_{rs}^{pq}\!=\!\!\int\!\!d\mathbf{x}_{1}d\mathbf{x}_{2}\phi_{p}^{*}(\mathbf{x}_{1})\phi_{q}^{*}(\mathbf{x}_{2})\frac{1}{\left|\mathbf{r}_{1}\!-\!\mathbf{r}_{2}\right|}\phi_{r}(\mathbf{x}_{2})\phi_{s}(\mathbf{x}_{1}).
\end{equation}
In these expressions, $Z_{I}$, $\mathbf{R}_{I}$ represent the atomic
number and position of the $I$-th nucleus, respectively, while $\mathbf{r}_{i}$
denotes the position of the $i$-th electron. The combined spatial
and spin coordinate of the $i$-th electron is denoted by $\textbf{x}_{i}=(\textbf{r}_{i},\sigma_{i})$,
where $\sigma_{i}$ represents the spin coordinate. The term $\frac{1}{\left|\mathbf{r}_{1}-\mathbf{r}_{2}\right|}$
describes the Coulomb repulsion between the two electrons.

Since the integration domain of electronic integrals spans the entire
spatial range, symmetry transformations do not alter the value of
the integrals. Mathematically, this invariance is expressed as:
\begin{equation}
h_{q}^{p}=\hat{\boldsymbol{O}}_{R}h_{q}^{p},\quad h_{rs}^{pq}=\hat{\boldsymbol{O}}_{R}h_{rs}^{pq}.
\end{equation}

Applying the symmetry transformation operator $\hat{\boldsymbol{O}}_{R}$
to the one-electron integral $h_{q}^{p}$, we derive:
\begin{equation}
\small
\begin{aligned}\hat{\boldsymbol{O}}_{R}h_{q}^{p} &\!=\!\!\int\!\!d\textbf{x}\hat{\boldsymbol{O}}_{R}\!\left[\!\phi_{p}^{*}(\mathbf{x})\!\left(\!\!-\frac{\nabla^{2}}{2}\!-\!\!\sum_{I}\!\frac{Z_{I}}{\left|\mathbf{r}\!-\!\mathbf{R}_{I}\right|}\!\right)\!\!\phi_{q}(\mathbf{x})\!\right]\\
=\int\!\!d\textbf{x} &\hat{\boldsymbol{O}}_{R}\phi_{p}^{*}(\mathbf{x})\!\!\left[\!\hat{\boldsymbol{O}}_{R}\!\!\left(\!-\frac{\nabla^{2}}{2}\!-\!\!\sum_{I}\!\!\frac{Z_{I}}{\left|\mathbf{r}\!-\!\mathbf{R}_{I}\right|}\right)\!\!\hat{\boldsymbol{O}}_{R}^{\dagger}\right]\!\!\hat{\boldsymbol{O}}_{R}\phi_{q}(\mathbf{x})\\
=\!\gamma_{p}^{R}\!\gamma_{q}^{R}\!&\! \int\! d\mathbf{x}\phi_{p}^{*}(\mathbf{x})\left(-\frac{\nabla^{2}}{2}-\sum_{I}\frac{Z_{I}}{\left|\mathbf{r}-\mathbf{R}_{I}\right|}\right)\phi_{q}(\mathbf{x})\\
\ =\!\gamma_{p}^{R}\!\gamma_{q}^{R}\!&h_{q}^{p}.
\end{aligned}
\end{equation}
In the second step, we use the invariance of the one-electron Hamiltonian
under symmetry transformations:
\begin{equation}
\small
\hat{\boldsymbol{O}}_{R}\!\left(\!-\frac{\nabla^{2}}{2}-\sum_{I}\frac{Z_{I}}{\left|\mathbf{r}\!-\!\mathbf{R}_{I}\right|}\right)\!\hat{\boldsymbol{O}}_{R}^{\dagger}\!=\!-\frac{\nabla^{2}}{2}-\sum_{I}\!\frac{Z_{I}}{\left|\mathbf{r}\!-\!\mathbf{R}_{I}\right|}.
\end{equation}

Combining the invariance condition $h_{q}^{p}=\hat{\boldsymbol{O}}_{R}h_{q}^{p}$
with the derived expression yields $h_{q}^{p}=\gamma_{p}^{R}\gamma_{q}^{R}h_{q}^{p}.$
If $\phi_{p}$ and $\phi_{q}$ belong to different irreps, there exists
at least one symmetry operation $R$ such that $\gamma_{p}^{R}\gamma_{q}^{R}=-1$.
Substituting this into the equation gives $h_{q}^{p}=-h_{q}^{p}$,
which directly implies $h_{q}^{p}=0$.

Following a similar derivation, for the two-electron integral $h_{rs}^{pq}$:
\begin{equation}
\small
\begin{aligned}\hat{\boldsymbol{O}}_{R}h_{rs}^{pq}\!=\!\!\!\int&\!\!d\mathbf{x}_{1}d\mathbf{x}_{2}\hat{\boldsymbol{O}}_{R}\!\!\left[\!\phi_{p}^{*}(\mathbf{x}_{1})\phi_{q}^{*}(\mathbf{x}_{2})\frac{1}{\left|\mathbf{r}_{1}\!-\!\mathbf{r}_{2}\right|}\phi_{r}(\mathbf{x}_{2})\phi_{s}(\mathbf{x}_{1})\!\right]\\
 \!=\!\int\!d\mathbf{x}_{1}d\mathbf{x}_{2}&\left(\hat{\boldsymbol{O}}_{R}\phi_{p}^{*}(\mathbf{x}_{1})\right)\!\!\left(\hat{\boldsymbol{O}}_{R}\phi_{q}^{*}(\mathbf{x}_{2})\right)\\
 \quad\!\!\cdot&\left(\!\hat{\boldsymbol{O}}_{R}\frac{1}{\left|\mathbf{r}_{1}\!-\!\mathbf{r}_{2}\right|}\hat{\boldsymbol{O}}_{R}^{\dagger}\!\right)\!\!\left(\!\hat{\boldsymbol{O}}_{R}\phi_{r}(\mathbf{x}_{2})\!\right)\!\!\left(\!\hat{\boldsymbol{O}}_{R}\phi_{s}(\mathbf{x}_{1})\!\right)\\
 \!=\!\gamma_{p}^{R}\!\gamma_{q}^{R}\!\gamma_{r}^{R}\!\gamma_{s}^{R}\!&\!\!\int\!\!d\mathbf{x}_{1}d\mathbf{x}_{2}\phi_{p}^{*}(\mathbf{x}_{1})\phi_{q}^{*}(\mathbf{x}_{2})\frac{1}{\left|\mathbf{r}_{1}\!-\!\mathbf{r}_{2}\right|}\phi_{r}(\mathbf{x}_{2})\phi_{s}(\mathbf{x}_{1})\\
 \!=\!\gamma_{p}^{R}\!\gamma_{q}^{R}\!\gamma_{r}^{R}\!\gamma_{s}^{R}\!&\ \!h_{rs}^{pq}.
\end{aligned}
\end{equation}
Here, we utilize the invariance of the Coulomb repulsion term under
spatial symmetry transformations:
\begin{equation}
\hat{\boldsymbol{O}}_{R}\frac{1}{\left|\mathbf{r}_{1}\!-\!\mathbf{r}_{2}\right|}\hat{\boldsymbol{O}}_{R}^{\dagger}=\frac{1}{\left|\mathbf{r}_{1}\!-\!\mathbf{r}_{2}\right|}.
\end{equation}

From the invariance condition $h_{rs}^{pq}=\hat{\boldsymbol{O}}_{R}h_{rs}^{pq}$,
we obtain $h_{rs}^{pq}=\gamma_{p}^{R}\gamma_{q}^{R}\gamma_{r}^{R}\gamma_{s}^{R}h_{rs}^{pq}$.
If the direct products of orbitals $\phi_{p}\otimes\phi_{q}$ and $\phi_{r}\otimes\phi_{s}$
belong to different irreps, there exists at least one symmetry operation
$R$ such that $\gamma_{p}^{R}\gamma_{q}^{R}\gamma_{r}^{R}\gamma_{s}^{R}=-1$.
This leads to $h_{pq}^{rs}=-h_{pq}^{rs}$, hence $h_{pq}^{rs}=0$.

In summary, we can draw the conclusion that, within an Abelian point
group, the electronic integrals in the Hamiltonian will vanish if
the corresponding excitation operators violate the point group symmetry
requirements. 
\end{proof}

Based on the above established theorem, we propose the HiUCCSD method,
whose pseudocode is presented in Algorithm \ref{algo}. Specifically,
HiUCCSD systematically prunes redundant excitation operators from
the input coupled-cluster excitation operator $\hat{T}$, by filtering
out those excitation operators linked to zero-valued coefficients
(i.e., electronic integrals) in the Hamiltonian. The reduced form
of $\hat{T}$ is then employed to construct the ansatz operator $e^{\hat{T}-\hat{T^{\dagger}}}$.

\textit{Notably, unlike SymUCCSD, the HiUCCSD ansatz does not require the
consideration of molecular structure and point group during implementation,
but instead only leverages the intrinsic information of the Hamiltonian.
Since the Hamiltonian is already calculated in the early steps of
the VQE algorithm implementation, this algorithm incurs almost no
additional computational cost.}

\section{Numerical Results}\label{sec:Numerical-Results}
\begin{table*}[ht]
    \centering
    \caption{Summary of molecular properties for VQE calculations, including the qubit count required, bond type, equilibrium bond length, and calculated electronic correlation energy. }
     \label{tab:moles}
    \setlength{\tabcolsep}{8pt}
    \begin{tabular}{ccccccc}
        \toprule  
        Molecule & Qubits  & \makecell{Bond\\Type} & \makecell{Equilibrium Bond\\Length ($\mathrm{\AA}$)} & \makecell{Correlation\\Energy} \\
        \midrule  
        HF              & 12       & H-F    & 1.00   & $3.27\times10^{-2}$ \\
        LiH             & 12       & Li-H   & 1.55   & $1.97\times10^{-2}$ \\
        $\mathrm{H_2O}$ & 14       & O-H    & 1.02   & $5.81\times10^{-2}$ \\
        $\mathrm{BeH_2}$& 14       & Be-H   & 1.32   & $3.46\times10^{-2}$ \\
        $\mathrm{NH_3}$ & 16       & N-H    & 1.09   & $7.30\times10^{-2}$ \\
        $\mathrm{CH_4}$ & 18       & C-H    & 1.11   & $8.25\times10^{-2}$ \\
        $\mathrm{N_2}$  & 20       & N-N    & 1.20   & $1.90\times10^{-1}$ \\
        CO              & 20       & C-O    & 1.20   & $1.46\times10^{-1}$ \\
        NaH             & 20       & Na-H   & 1.64   & $5.55\times10^{-2}$ \\
        $\mathrm{C_2H_4}$ & 28     & C-C    & 1.24   & $1.63\times10^{-1}$ \\
        \bottomrule  
    \end{tabular}
\end{table*}
\begin{table*}[ht]
    \centering
    \caption{VQE simulation results for molecules at their equilibrium geometries. Columns 2–3 report the molecular point group and whether it is Abelian, respectively. Columns 4–6 list the parameter counts of three ansatzes (SymUCCSD, HiUCCSD, UCCSD), where the value in parentheses indicates the Abelian subgroup utilized in its implementation. Row 7 presents the parameter reduction ratio of HiUCCSD relative to UCCSD. The final three columns show the energy deviations of each method with respect to the exact FCI result. VQE results showing significant precision deficiency are labeled with \bcancel{strikethrough}.}
    \label{tab:vqe_res}
    \small
    \setlength{\tabcolsep}{5pt}
    \begin{tabular}{cccccccccc}
        \toprule 
        Molecule & \makecell{Point\\Group} & Abelian  & SymUCCSD & HiUCCSD & UCCSD & \makecell{HiUCCSD\\Reduction} & $\Delta E_{\text{SymUCCSD}}$ & $\Delta E_{\text{HiUCCSD}}$ & $\Delta E_{\text{UCCSD}}$ \\
        \midrule 
        HF               & $C_{\infty\text{v}}$   & No  & 11 ($C_{2\text{v}}$) & 11   & 20   & $45\%$  & $8.82\times10^{-7}$          & $8.82\times10^{-7}$  & $8.82\times10^{-7}$\\
        LiH              & $C_{\infty\text{v}}$   & No  & 20 ($C_{2\text{v}}$) & 20   & 44   & $55\%$  & $1.02\times10^{-5}$          & $1.02\times10^{-5}$  & $1.02\times10^{-5}$ \\
        $\mathrm{H_2O}$  & $C_{2\text{v}}$        & Yes & 26 ($C_{2\text{v}}$) & 26   & 65   & $60\%$  & $1.21\times10^{-4}$          & $1.21\times10^{-4}$  & $1.21\times10^{-4}$ \\
        $\mathrm{BeH_2}$ & $D_{\infty\text{h}}$   & No  & 23 ($D_{2\text{h}}$) & 23   & 90   & $74\%$  & $3.77\times10^{-4}$           & $3.77\times10^{-4}$ & $3.77\times10^{-4}$ \\
        $\mathrm{NH_3}$  & $C_{3\text{v}}$        & No  & 75 ($C_{\text{s}}$)  & 93   & 135  & $32\%$  & $\bcancel{2.78\times10^{-2}}$ & $1.09\times10^{-4}$ & $1.09\times10^{-4}$ \\
        $\mathrm{CH_4}$  & $T_{\text{d}}$         & No  & 65 ($D_{2}$)         & 188  & 230  & $18\%$  & $\bcancel{4.08\times10^{-2}}$ & $2.17\times10^{-4}$ & $2.17\times10^{-4}$ \\
        $\mathrm{N_2}$   & $D_{\infty\text{h}}$   & No  & 49 ($D_{2\text{h}}$) & 65   & 252  & $74\%$  & $\bcancel{4.02\times10^{-2}}$ & $2.80\times10^{-3}$ & $2.80\times10^{-3}$ \\
        CO               & $C_{\infty\text{v}}$   & No  & 80 ($C_{2\text{v}}$) & 106  & 252  & $58\%$  & $\bcancel{1.24\times10^{-2}}$ & $9.32\times10^{-3}$ & $9.32\times10^{-3}$ \\
        NaH              & $C_{\infty\text{v}}$   & No  & 110 ($C_{2\text{v}}$)& 148  & 324  & $54\%$  & $\bcancel{2.09\times10^{-2}}$ & $2.09\times10^{-3}$ & $2.09\times10^{-3}$ \\
        $\mathrm{C_2H_4}$& $D_{2\text{h}}$        & Yes & 219 ($D_{2\text{h}}$)& 219  & 1224 & $83\%$  & $8.28\times10^{-3}$ & $8.28\times10^{-3}$ & $8.28\times10^{-3}$ \\
        \bottomrule
    \end{tabular}
\end{table*}

In this section, PySCF \cite{pyscf} is utilized to calculate electronic integrals and identify the point groups of molecules, while MindSpore Quantum \cite{MindQuantum} is employed to perform second quantization, Jordan-Wigner transformation, and quantum circuit simulations. The qubit-excitations scheme \cite{qubit-excitation-operator,efficient_excitations_circ} is adopted for implementing fermionic excitation operators. All calculations are performed with the minimal STO-3G basis set without considering frozen orbitals, and all optimization procedures employ the Broyden-Fletcher-Goldfarb-Shanno (BFGS) algorithm \cite{bfgs} in the SciPy Python package \cite{scipy}. The UCCSD ansatz is constructed using spin-complemented single- and double-excitation operators, and all simulations in this study are conducted under idealized conditions where both sampling noise and hardware-induced noise are neglected. The code and data supporting this study are publicly available at \href{https://atomgit.com/mindspore/mindquantum/tree/research/paper_with_code/HiUCCSD}{https://atomgit.com/mindspore/mindquantum/tree} \href{https://atomgit.com/mindspore/mindquantum/tree/research/paper_with_code/HiUCCSD}{/research/paper\_with\_code/HiUCCSD}.

\subsection{VQE Implementation with Reduced Ansatz }\label{subsec:VQE-Implementation}

\begin{figure*}[ht]
\centering
\includegraphics[scale=0.65]{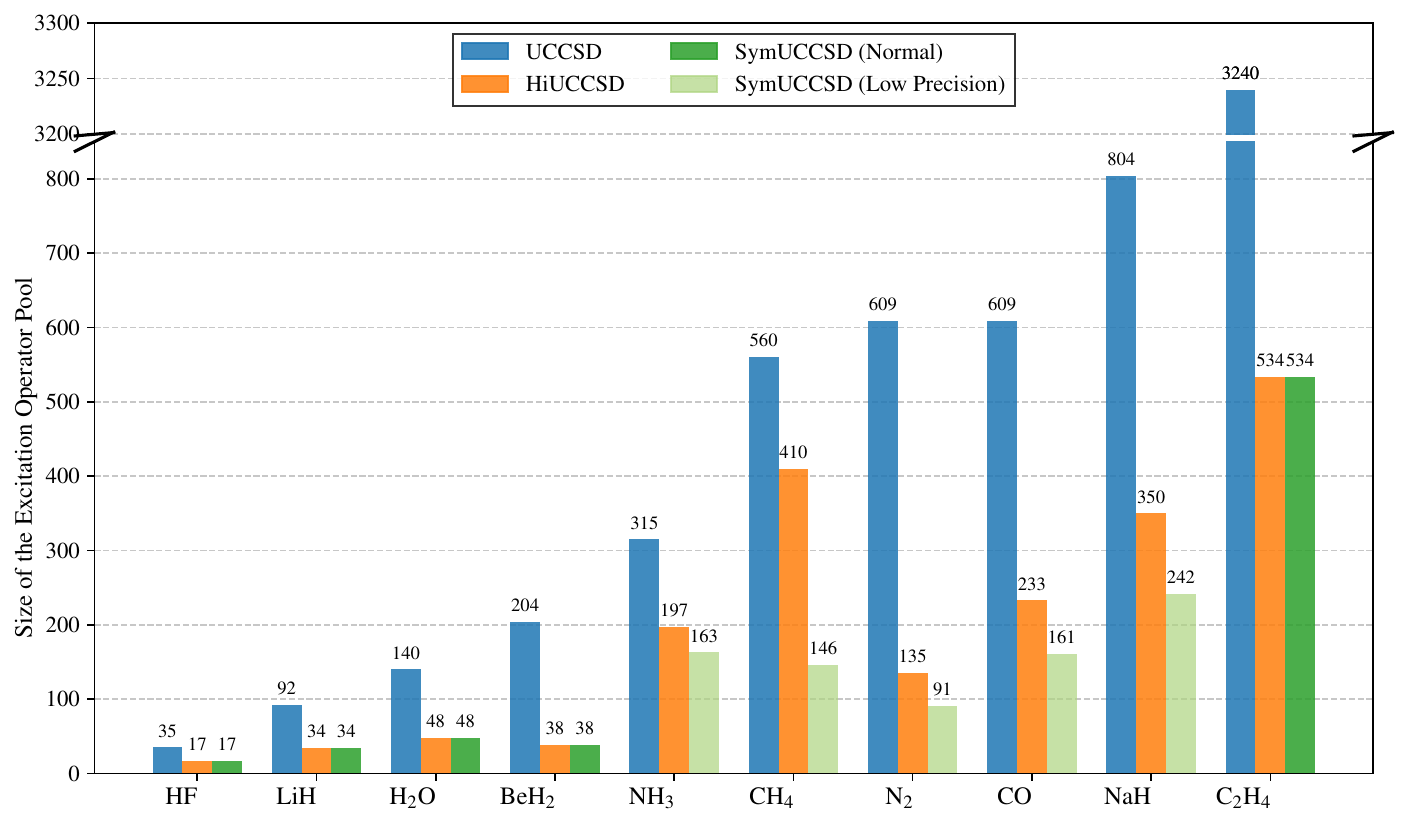}
\caption{Size of excitation operator pools generated by SymUCCSD, HiUCCSD and
UCCSD for various molecules. SymUCCSD cases that yield low-precision
VQE results are labeled in light-green.}
\label{fig:pool}
\end{figure*}

To evaluate the performance of the UCCSD, SymUCCSD, and HiUCCSD, in
this subsection, we run the standard VQE algorithm based on these
three ansatzes to calculate the ground-state energies of a variety
of small molecules. These molecules cover several common Abelian and
non-Abelian point groups.

Table \ref{tab:moles} summarizes the properties of ten representative
small molecules for VQE calculations. Specifically: “Qubits” denotes
the number of qubits required in simulations (reflecting the computational
scale), “Equilibrium
Bond Length” (for the specified “Bond Type”) is derived from geometric
optimization, and the “Correlation Energy” is calculated as the absolute
difference between the exact FCI energy and the Hartree-Fock energy. 

Table \ref{tab:vqe_res} presents the VQE simulation results of the above molecules at their equilibrium geometries. The table first lists each molecule’s highest-symmetry point group, with an additional label “Abelian” indicating whether the group is Abelian. Rows 4–6 report the parameter counts of each ansatz. Row 7 presents the percentage reduction in the number of parameters for HiUCCSD relative to UCCSD. The last three columns  ($\Delta E_{\text{SymUCCSD}}$, $\Delta E_{\text{HiUCCSD}}$,
$\Delta E_{\text{UCCSD}}$) quantify the energy deviations of each
method relative to the exact FCI result. All parameters are initialized
to 0 in the VQE simulations to eliminate the potential influence of
parameter initialization on the calculation results. We note that
SymUCCSD can only be directly implemented for molecules belonging
to Abelian point groups \cite{VQE_point_group_symmetry}. Thus, in
practice, the molecular point group is reduced to its highest-order
Abelian subgroup to enable the implementation of SymUCCSD. The Abelian
subgroups adopted are explicitly indicated in parentheses following
the parameter counts of SymUCCSD (in the fourth column of Table \ref{tab:vqe_res}).
For instance, NH$_{3}$ has an actual point group of $C_{3\text{v}}$
(a 6-order non-Abelian group), yet it is restricted to the $C_{\text{s}}$
group (a 2-order Abelian group) in SymUCCSD.

As shown in Table \ref{tab:vqe_res}, for all molecules belonging
to Abelian point groups (H$_{2}$O, C$_{2}$H$_{4}$) as well as certain
molecules belonging to non-Abelian point groups (HF, LiH, BeH$_{2}$),
both SymUCCSD and HiUCCSD attain equivalent parameter reduction performance
compared to UCCSD while preserving high computational accuracy. This
observation validates the equivalence of the two methods for molecules
with Abelian point groups, as theoretically proven in Section \ref{subsec:HiUCCSD}
above.

Nevertheless, for some molecules belonging to non-Abelian point groups
(NH$_{3}$, CH$_{4}$, N$_{2}$, CO, NaH), SymUCCSD exhibits distinctly
insufficient computational accuracy. This is attributed to the omission
of critical excitation operators, which leads to limited expressiveness
of the ansatz. As a result, SymUCCSD may not be applicable to molecules
with non-Abelian point groups, even when Abelian subgroups are adopted
in its implementation. In contrast, HiUCCSD not only achieves significant
parameter simplification but also preserves computational accuracy
that is consistent with that of the original UCCSD across all molecular
systems investigated. For the molecules investigated, the HiUCCSD
ansatz reduces the number of parameters by $18\%$ to $83\%$ relative
to the UCCSD ansatz, without compromising the calculation accuracy.

\subsection{ADAPT-VQE Implementation with Reduced Operator Pools }\label{subsec:ADAPT-VQE-Implementation}

\begin{figure*}[htbp] 
    \centering
    \includegraphics[scale=0.68]{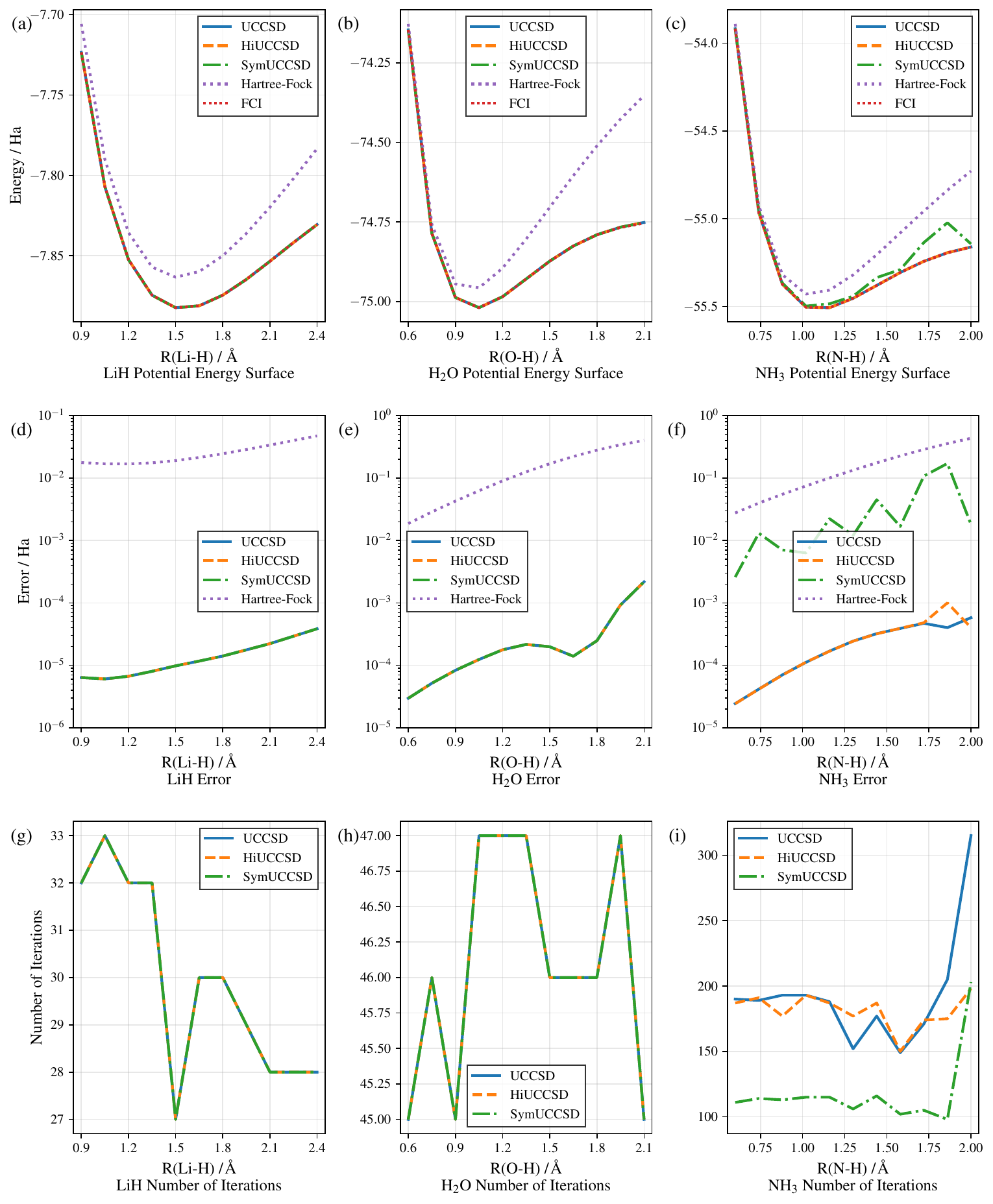}
    \caption{Potential energy surfaces (a--c), solution errors (d--f), and number of iterations (g--i) versus bond length obtained by ADAPT-VQE with different operator pools for the $\mathrm{LiH}$, $\mathrm{H_2O}$, and $\mathrm{NH_3}$ molecular systems.}
    \label{fig:adapt}
\end{figure*}

In the previous subsection, we discussed the VQE calculations of SymUCCSD,
HiUCCSD, and standard UCCSD under equilibrium geometries. Here, we
discuss the performance of these three methods in the ADAPT-VQE algorithm.

As shown in Fig.~\ref{fig:pool}, we compare the sizes of excitation
operator pools generated by three VQE ansatzes across ten target molecules.
For all Abelian systems (H$_{2}$O, C$_{2}$H$_{4}$) and certain
non-Abelian systems (HF, LiH, BeH$_{2}$), both SymUCCSD and HiUCCSD
achieve a substantial reduction in operator pool size relative to
UCCSD: for instance, the pool size of UCCSD reaches 3240 for C$_{2}$H$_{4}$,
whereas that of SymUCCSD (normal case) and HiUCCSD is only 534. Notably,
the light-green bars in the figure denote SymUCCSD cases that yield
low-precision VQE results for specific non-Abelian molecules (NH$_{3}$,
CH$_{4}$, N$_{2}$, CO, NaH), which facilitates distinguishing the
performance reliability of SymUCCSD across different molecular systems.
These results demonstrate that HiUCCSD maintains a compact operator
pool without compromising calculation precision, while SymUCCSD may
suffer precision loss in some cases despite its small pool size. Overall,
for all investigated molecules, HiUCCSD reduces the size of the excitation
operator pool by 27$\%$ to 84$\%$ relative to UCCSD. This represents
a significant reduction in computational cost for ADAPT-VQE implementation,
as selecting the optimal operator in each iteration constitutes a
key bottleneck of the algorithm.

Next, to provide a concrete illustration, we investigate the ground-state
energy calculations of three representative molecules (LiH, H$_{2}$O
and NH$_{3}$) under bond stretching, where the electron correlation
effects are significant. As shown in Table \ref{tab:vqe_res}, H$_{2}$O
belongs to an Abelian point group ($C_{2\text{v}}$), while LiH and
NH$_{3}$ belong to non-Abelian point groups ($C_{\infty\text{v}}$
and $C_{3\text{v}}$).
The ADAPT-VQE algorithm is configured here to terminate the iteration when either the norm of the gradient vector of the operator pool drops below 
$10^{-3}$ or the number of iterations exceeds the size of the operator pool. Each excitation operator in the operator pool uses independent
parameters, which ensures stable iterative convergence when the warm-start
training strategy is adopted \cite{ADAPT-VQE}.

Fig.~\ref{fig:adapt} presents a visual comparison of the potential
energy surfaces, solution errors relative to FCI results, and iteration
counts derived from ADAPT-VQE implementations with distinct operator
pools, for three target molecular systems (LiH, H$_{2}$O, and NH$_{3}$).
As illustrated, for H$_{2}$O (an Abelian point group molecule) and
LiH (a non-Abelian point group molecule), all three operator pools
yield sufficiently high accuracy. In contrast, for NH$_{3}$ (a non-Abelian
point group molecule), only the HiUCCSD pool achieves performance
comparable to that of the UCCSD pool, whereas the SymUCCSD pool fails
to deliver satisfactory results. This observation indicates that the
SymUCCSD pool omits certain critical excitation operators for NH$_{3}$.
This deficiency impairs the ansatz’s expressive power, thereby preventing
it from adequately approximating the true potential energy surface. 

Based on the numerical results in this section, we can draw the following
conclusions: for molecules belonging to the Abelian point group, both
SymUCCSD and HiUCCSD can achieve significant simplification while
retaining the expressive power of the full UCCSD ansatz. However,
for molecules belonging to the non-Abelian point group, the
expressive power of SymUCCSD may decrease, while HiUCCSD remains
unaffected. We note that for applications involving non-Abelian point
groups, the theoretical underpinnings of our method have not yet been
fully established at this stage, and relevant conclusions are currently
supported by numerical results; we intend to further elaborate on
this aspect in future work. Even so, our results still indicate that
HiUCCSD is more robust and easier to implement than SymUCCSD.

\section{Conclusion}\label{sec:Conclusion}
In this paper, we propose a novel method for constructing point group-respecting ansatz rooted in the intrinsic information of the Hamiltonian,
termed HiUCCSD. We have theoretically proven the effectiveness of
this method for molecules belonging to Abelian point groups. We conducted
a performance comparison between the HiUCCSD and SymUCCSD
ansatzes via numerical experiments implementing the VQE and ADAPT-VQE
algorithms across ten small molecular systems. The experimental results
indicate that HiUCCSD and SymUCCSD exhibit comparable performance
for molecules belonging to Abelian point groups and a subset of non-Abelian
point group molecules. In contrast, for certain non-Abelian point
group molecules, SymUCCSD suffers from performance degradation, whereas
HiUCCSD maintains stable performance while achieving effective ansatz
simplification. Overall, for all investigated molecules, HiUCCSD reduces
the number of parameters for VQE and the size of the excitation operator
pool for ADAPT-VQE by 18$\%$--83$\%$ and 27$\%$--84$\%$,
respectively, relative to UCCSD. These reductions correspond to significant
reductions in computational cost and circuit depth. These results
empirically verify that HiUCCSD exhibits enhanced robustness and a
broader scope of applicability, presenting a novel symmetry-respecting
ansatz candidate to facilitate the realization of practical quantum
advantages in quantum chemical simulations on NISQ devices.

\bibliography{References_library}
\end{document}